\documentclass[twocolumn,showpacs,aps,prd]{revtex4}

\usepackage{graphicx}
\usepackage{dcolumn}
\usepackage{amsmath}
\usepackage{epsfig}

\input pubboard/babarsym

\newcommand{\BABARPubYear}    {10}
\newcommand{\BABARPubNumber}  {017}

\newcommand{\SLACPubNumber} {14270}

\def\figurebox#1#2#3{%
    \def\arg{#3}%
    \ifx\arg\empty
    {\hfill\vbox{\hsize#2\hrule\hbox to #2{\vrule\hfill\vbox to #1{\hsize#2\vfill}\vrule}\hrule}\hfill}%
    \else
    {\hfill\epsfbox{#3}\hfill}%
    \fi}

\def\bgg      {\ensuremath{\B^{0} \to \gamma \gamma}}

\def\LRpiz    {\ensuremath{L_{\piz}}}
\def\LReta    {\ensuremath{L_{\eta}}}
\def\result   {\ensuremath{3.3\times10^{-7}}}

\begin{document}

\preprint{\babar-PUB-\BABARPubYear/\BABARPubNumber} 
\preprint{SLAC-PUB-\SLACPubNumber} 

\begin{flushleft}
  \babar-PUB-\BABARPubYear/\BABARPubNumber\\
  SLAC-PUB-\SLACPubNumber\\
\end{flushleft}

\title{
{\large \bf
Search for the Decay \boldmath{\bgg}} 
}

%
\author{P.~del~Amo~Sanchez}
\author{J.~P.~Lees}
\author{V.~Poireau}
\author{E.~Prencipe}
\author{V.~Tisserand}
\affiliation{Laboratoire d'Annecy-le-Vieux de Physique des Particules (LAPP), Universit\'e de Savoie, CNRS/IN2P3,  F-74941 Annecy-Le-Vieux, France}
\author{J.~Garra~Tico}
\author{E.~Grauges}
\affiliation{Universitat de Barcelona, Facultat de Fisica, Departament ECM, E-08028 Barcelona, Spain }
\author{M.~Martinelli$^{ab}$}
\author{A.~Palano$^{ab}$ }
\author{M.~Pappagallo$^{ab}$ }
\affiliation{INFN Sezione di Bari$^{a}$; Dipartimento di Fisica, Universit\`a di Bari$^{b}$, I-70126 Bari, Italy }
% ...
\collaboration{The \babar\ Collaboration}
\noaffiliation

\date{\today}

\begin{abstract}
We report the result  of a search for the rare decay  \bgg\ in $426\, \invfb$ of
data,  corresponding  to 226  million  \BzBzb  pairs,  collected on  the  \FourS
resonance  at  the \pep2  asymmetric-energy  \epem  collider  using the  \babar\
detector.   We use  a maximum  likelihood fit  to extract  the signal  yield and
observe $21^{+13}_{-12}$ signal events with a statistical signficance of $1.9\,
\sigma$.  This corresponds to a branching fraction ${\cal B}(\bgg) = (1.7 \pm 
1.1  {\rm (stat.)}   \pm 0.2  {\rm (syst.))}   \times 10^{-7}$.   Based  on this
result,  we  set a  90\%  confidence  level upper  limit  of  ${\cal B}(\bgg)  <
\result$.

\end{abstract}

\pacs{13.20.He}

\maketitle

\section{Introduction}
\label{sec:intro}
\begin{figure}

  \includegraphics[width=\linewidth,clip=true]{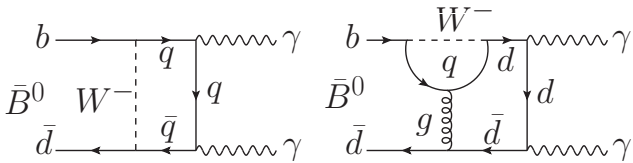}

  \caption
  {
    \label{fig:diagrams}
    Examples  of lowest  order SM  Feynman diagrams  for \bgg.   The  symbol $q$
    represents a  $u$, $c$, or  $t$ quark.  In  some new physics  scenarios, the
    $W$-boson may be replaced by a charged Higgs particle.  
  }

\end{figure}

In the standard  model (SM), the decay $\Bz \to \gamma  \gamma$ occurs through a
flavor-changing  neutral current  (FCNC) transition  involving  electroweak loop
diagrams,  as illustrated  in  Figure~\ref{fig:diagrams}.  The  decays $\Bz  \to
\gamma \gamma$ and $B_s \to \gamma  \gamma$ are closely related, with the $b \to
d \gamma  \gamma$ transition being  suppressed with respect  to $b \to  s \gamma
\gamma$ by Cabbibo-Kobayashi-Maskawa  (CKM) factors ($|V_{td}|^2/|V_{ts}|^2 \sim
0.04$).   Hadron  dynamics  introduces  uncertainties  into  the  prediction  of
branching fractions  for these  decays and  may modify the  ratio away  from the
CKM-implied value.  While $\Bz \to \gamma  \gamma$ is expected to have a smaller
branching fraction than  $B_s \to \gamma \gamma$, a search  for the latter faces
the experimental challenge of obtaining  a large sample of $B_s$ mesons, whereas
large samples of \Bz mesons  are readily available from \BF\ experiments running
on the $\Upsilon(4S)$ resonance.

A  leading order  calculation  for the  branching  fraction of  \bgg\ yields  an
estimate  of   $3.1^{+6.4}_{-1.6}  \times  10^{-8}$~\cite{BB}.    This  mode  is
sensitive to  new physics  that could  lead to an  enhancement of  the branching
fraction due  to possible contributions  of non-SM heavy particles  occurring in
the  loop of  the  leading-order  Feynman diagrams.   Such  enhancements to  the
branching fraction for \bgg\ are less constrained than those for $B_s \to \gamma
\gamma$ due  to the fact that the  $b \to s \gamma$  transition, responsible for
$B_s \to  \gamma \gamma$, is known much  more accurately than $b  \to d \gamma$.
For example, some  new physics scenarios involving an  extended Higgs sector may
considerably   enhance  the  branching   fractions  with   respect  to   the  SM
expectation~\cite{Aliev-Iltan}.          Supersymmetry        with        broken
$R$-parity~\cite{Gemintern-etal}   also  provides   scenarios  where   order  of
magnitude enhancements  are possible.  In  addition, since the  two-photon final
state can be either $\CP$-even  or $\CP$-odd, studies of $\CP$-violating effects
may ultimately be possible.

The best previous upper limit on the branching fraction at 90\% confidence level
(CL) is ${\cal B}(\Bz \to \gamma \gamma)  < 6.2 \times 10^{-7}$ set by the Belle
experiment~\cite{Belle-Villa} using  a dataset recorded at  the \FourS resonance
with an integrated luminosity of $104  \, \invfb$.  For the related process $B_s
\to \gamma \gamma$, Belle has set a  an upper limit on the branching fraction of
$8.7 \times 10^{-6}$ (90\% CL) based on $23.6 \, {\rm fb^{-1}}$ of data taken on
the $\Upsilon(5S)$ resonance~\cite{Belle-Wicht}.

We report herein a new search for  the decay \bgg\ which uses a data sample with
integrated luminosity  of $426\,  \invfb$ taken at  the \FourS  resonance.  This
corresponds to the entire \babar\ \FourS dataset and contains 226 million \BzBzb
pairs.  The analysis  does not distinguish between \Bz  and \Bzb, and throughout
this article, charge conjugation is implied for all reactions.

The analysis proceeds through several  steps.  The full dataset is first reduced
to  a manageable  size by  selecting events  based on  loose  kinematic criteria
consistent with the \bgg\ hypothesis. Studies are then performed to determine an
optimal set of event selection criteria to maximize the efficiency for detecting
\bgg\ events  while effectively rejecting  background events.  This  analysis is
performed  ``blind''  in  the  sense  that  the  event  selection  criteria  are
determined without  considering the on-resonance  data within a  specific signal
region, as defined below.  We use an unbinned extended maximum likelihood fit to
extract  the  signal  yield  from  the  remaining  events.   Finally,  since  no
statistically  significant \bgg\  signal  is  observed, an  upper  limit on  the
branching  fraction is calculated  based on  the likelihood  function determined
from the fit.

Section~II  of  this article  describes  the  \babar\  detector and  the  \FourS
dataset.  Section~III outlines the  optimization of the event selection criteria
and discusses backgrounds due to  exclusive \B decays.  Section~IV describes the
fit methodology and Section~V discusses the sources of systematic uncertainties.
Finally,  Section~VI reports  the  resulting branching  fraction  and the  upper
limit.

\section{The \babar\ Detector and Dataset}
\label{sec:detector}
The  \babar\ detector  recorded data  from the  \pep2 \BF\  located at  the SLAC
National  Accelerator  Laboratory.   In  \pep2, head-on  collisions  of  9.0~GeV
electrons  and  3.1~GeV  positrons  provide  a  center-of-mass  (CM)  energy  of
10.58~GeV  that lies  at the  peak of  the \FourS  resonance.  The  \FourS meson
decays almost  exclusively to \BB pairs.   The subsequent $B$  meson decays were
observed  in  the   \babar\  detector,  which  has  been   described  in  detail
elsewhere~\cite{BABARNIM}.  Briefly, a superconducting solenoid produces a 1.5~T
magnetic  field approximately parallel  to the  colliding electron  and positron
beams.  Most detector subsystems are inside the volume of the solenoid.  Charged
particle tracking is accomplished by  a combination of a five layer double-sided
silicon  strip  vertex  detector  and  a  40 layer  drift  chamber.   The  track
reconstruction algorithm accepts tracks  with a transverse momentum greater than
$50\, \mevc$.   Identification  of  charged  particles  is  accomplished  using  a
ring-imaging Cherenkov  detector augmented with energy loss  measurements in the
tracking  detectors.  Photons  are detected  in the  electromagnetic calorimeter
(EMC), which  consists of 6580~CsI(Tl) crystals, oriented  in a quasi-projective
geometry with  respect to the interaction  point.  Outside of  the solenoid, the
steel flux return is instrumented with a combination of resistive-plate chambers
and limited streamer tubes to  provide detection of muons and long-lived neutral
hadrons.

The  EMC is  the most  important detector  subsystem for  the \bgg\  search.  It
provides  polar angle  coverage in  the  laboratory frame  from $15.8^\circ$  to
$141.8^\circ$,  and  full azimuthal  coverage,  corresponding  to a  solid-angle
coverage of 90\% in the $\Upsilon(4S)$ CM frame.
When  a photon  interacts  with the  EMC  it creates  an electromagnetic  shower
depositing  its  energy into  many  contiguous  crystals,  typically 10  to  30,
hereafter called a ``cluster''.  If no track in the event points to the cluster,
it is designated as a photon candidate.
Individual crystals are  read out by a pair of  silicon PIN photodiodes attached
to the  rear face of the  crystal.  Amplified signals are  sampled and digitized
with a  period of 270~ns, providing a  continuous data stream to  which gain and
pedestal corrections are applied.  When  a first level trigger is recorded, data
samples in  a time window of $\pm1  \, \mu$s are selected,  producing a waveform
which is  analyzed by  a feature  extraction algorithm running  in real  time in
readout modules.  For events passing a higher level trigger, any EMC signal with
energy  above  a  0.5~\mev  threshold   has  its  deposited  energy  and  timing
information recorded for offline analysis.

The energy resolution of the EMC is  parameterized as the sum of two terms added
in quadrature, given by~\cite{SLAC-EMC}:
\begin{equation*}\label{eq:eRes}
  \frac{\sigma_E}{E} = {2.30 \% \over \sqrt[4]{E\, (\gev)}} \oplus 1.35\%, 
\end{equation*}
while the angular resolution in the polar angle $\theta$ and the azimuthal angle
$\phi$ is given by:
\begin{equation*}\label{eq:angRes}
 \sigma_\theta = \sigma_\phi = {4.16 \over \sqrt{E\, (\gev)}} \, {\rm mrad}.
\end{equation*}

In addition to  the on-resonance data, a data sample of  $44$ \invfb taken about
40~\mev below the  \FourS peak is recorded and used to  validate the Monte Carlo
(MC) simulation of continuum processes, \epem\to\qqbar, where $q = u,d,s,\, {\rm
or}\ c)$ and  \epem\to\tautau.  MC simulated events are  produced using the {\tt
EVTGEN}~\cite{evtGen} package to  model the physics of \B  meson decays and {\tt
JETSET}~\cite{jetset}   to  model   quark  fragmentation.    The   {\tt  GEANT}4
toolkit~\cite{geant4} is used  to simulate the interaction of  these events with
the detector model.   These tools are designed to take  into account the varying
detector and beam conditions encountered during data-taking.  The MC events were
analyzed with  the same reconstruction  algorithms, event selection  and fitting
procedures as  data.  MC samples  of $B \Bbar$  events correspond to  about four
times  the  integrated  luminosity of  the  data.   Those  for  $e^+ e^-  \to  c
\overline{c}$ events  correspond to twice  the data luminosity, while  those for
$e^+ e^- \to q  \overline{q}$ (where $q = u, \, d, {\rm or}  \, s$) and $e^+ e^-
\to \tau^+ \tau^-$ correspond to  approximately the same luminosity as data.  In
addition, special  MC data sets  are created in  which large samples of  rare \B
meson decays  are generated  for the purpose  of investigating the  signal \bgg\
decay as well as possible backgrounds due to other \B decays.

\section{Event Selection And Backgrounds}
\label{sec:evtReco}

\subsection{Event Selection}
\label{sec:evtSel}
The full \FourS  on-resonance dataset is first reduced  by selecting events that
contain at least two photons with energies of $1.15 \leq E^*_{\gamma} \leq 3.50$
\gev,  where the  asterisk indicates  a  quantity in  the \FourS  CM frame.   We
consider all  combinations of two photons  whose energies lie in  this range and
add their four-momentum to create \B meson candidates.  Hereafter, these photons
are  referred  to  as  \B  candidate photons.   The  distribution  of  correctly
reconstructed \B candidates will peak in two nearly uncorrelated variables, \mes
and \DeltaE.  The beam energy substituted mass is defined as $\mes \equiv \sqrt{
E^{*2}_{\rm  beam} -  \vec{p}^{*2}_{\rm B}}/c^2$  and the  energy  difference is
$\DeltaE \equiv  E^*_{\rm B}  - E^*_{\rm beam}$,  where $E^*_{\rm beam}$  is the
beam energy,  and $\vec{p}^*_{\rm B}$  and $E^*_{\rm B}$ are  the three-momentum
and  energy of  the  \B candidate,  respectively.   For \bgg\  events, the  \mes
distribution will peak at the \B meson mass, $5.279$ \gevcc~\cite{PDG2010}.  The
MC predicts  a full-width at half  maximum (FWHM) of $6.5  \mevcc$.  The \DeltaE
distribution is asymmetric  and will peak near zero with a  tail to the negative
\DeltaE side  due to photon  energy loss outside  the active volume of  the EMC.
The FWHM  for \DeltaE predicted  by the  MC is about  $150 \mev$.  We  select an
event for further analysis if it  contains exactly one \B candidate with $\mes >
5.1$ \gevcc and $-0.50  \leq \DeltaE \leq 0.50$ \gev.  We find  that in \bgg\ MC
only $0.06\%$ of events have more than  one \B candidate and are removed by this
selection.    Events  from   continuum   processes,  $\epem   \to  \qqbar$   and
\epem\to\tautau, are suppressed  by requiring the ratio of  the second to zeroth
Fox-Wolfram moments~\cite{fox},  $R_2$, to be  less than $0.90$.  This  ratio is
calculated from the  momenta of all charged and neutral  particles in the event.
To suppress  backgrounds from \epem\to\tautau  events, which tend to  have lower
multiplicity  compared to  \bgg\  events, the  number  of reconstructed  charged
tracks in the event is required to be greater than two.

We define a signal region in  the \mes-\DeltaE plane by fitting each variable in
\bgg\ MC  and selecting a range around  the peak of the  distribution.  The \mes
distribution is fit  with a Crystal Ball (CB) shape~\cite{Gaiser}  and we take a
$\pm 3\, \sigma$ region around the CB peak corresponding to $5.27 \leq \mes \leq
5.29$ \gevcc.   The \DeltaE distribution  is parameterized using  a double-sided
modified Gaussian with tail parameters given by
\begin{equation}\label{eq:cruijff}
  f(\DeltaE) \propto \exp 
  \left(
    \frac{-(\DeltaE - \mu)^2}
	 {2\sigma^2_{\rm L,R} + \alpha_{\rm L,R}(\DeltaE - \mu)^2}
  \right),
\end{equation}
where $\mu$ is the peak of  the \DeltaE distribution, $\sigma_{\rm L,R}$ are the
distribution widths, and $\alpha_{\rm L,R}$  are tail parameters on the left and
right side of the peak, respectively.  We again take a $\pm 3\, \sigma$ interval
around  the peak corresponding  to $-0.30  \leq \DeltaE  \leq 0.13$  \gev.  This
region is blinded  in the on-resonance data until the  maximum likelihood fit is
performed.  With this  definition of the signal region,  an \mes sideband region
is defined as $5.20 < \mes < 5.27$ \gevcc, and lower and upper \DeltaE sidebands
are defined as $-0.50 \leq \DeltaE <  -0.30$ \gev and $0.13 < \DeltaE \leq 0.50$
\gev, respectively.

We  select photon  candidates  according  to criteria  chosen  to remove  poorly
reconstructed photons.  The  energy of the cluster must be  spread over at least
10  crystals with all  crystals in  the cluster  having active  electronics with
correct calibrations.  The shape of the  cluster must be consistent with that of
a photon  in the  defined energy  range; we therefore  require clusters  to have
lateral moments~\cite{LAT} in  the range $0.15 \leq f_{\rm  lat} \leq 0.50$.  To
ensure the shower  is fully contained within the EMC  volume, only photons whose
polar  angle is in  the range  $22.9^{\circ} <  p_{\theta} <  137.5^{\circ}$ are
selected.  The photons  are kept for further analysis if  they are isolated from
all other clusters in the event by at least 25 cm.

If an \epem collision that results  in a trigger is accompanied by another \epem
collision  nearby  in time,  EMC  signals  from  the out-of-time  collision  may
populate  the event  of  interest.   Due to  the  large Bhabha-scattering  cross
section, these ``pile-up'' events  often involve high-energy electrons which may
produce  EMC  clusters.  Since  the  tracking  detectors  are sensitive  over  a
narrower time window, electron-induced tracks may not point to the EMC clusters,
causing  them to  be treated  as  photon candidates.   If their  energy is  also
measured incorrectly, artificial \bgg\  candidates may result.  This scenario is
effectively  rejected  by requiring  the  total event  energy  to  be less  than
$15.0$~\gev and  the cluster time of  each \B candidate photon  to be consistent
with the trigger event time.

\begin{figure}

  \includegraphics[width=\linewidth,clip=true]{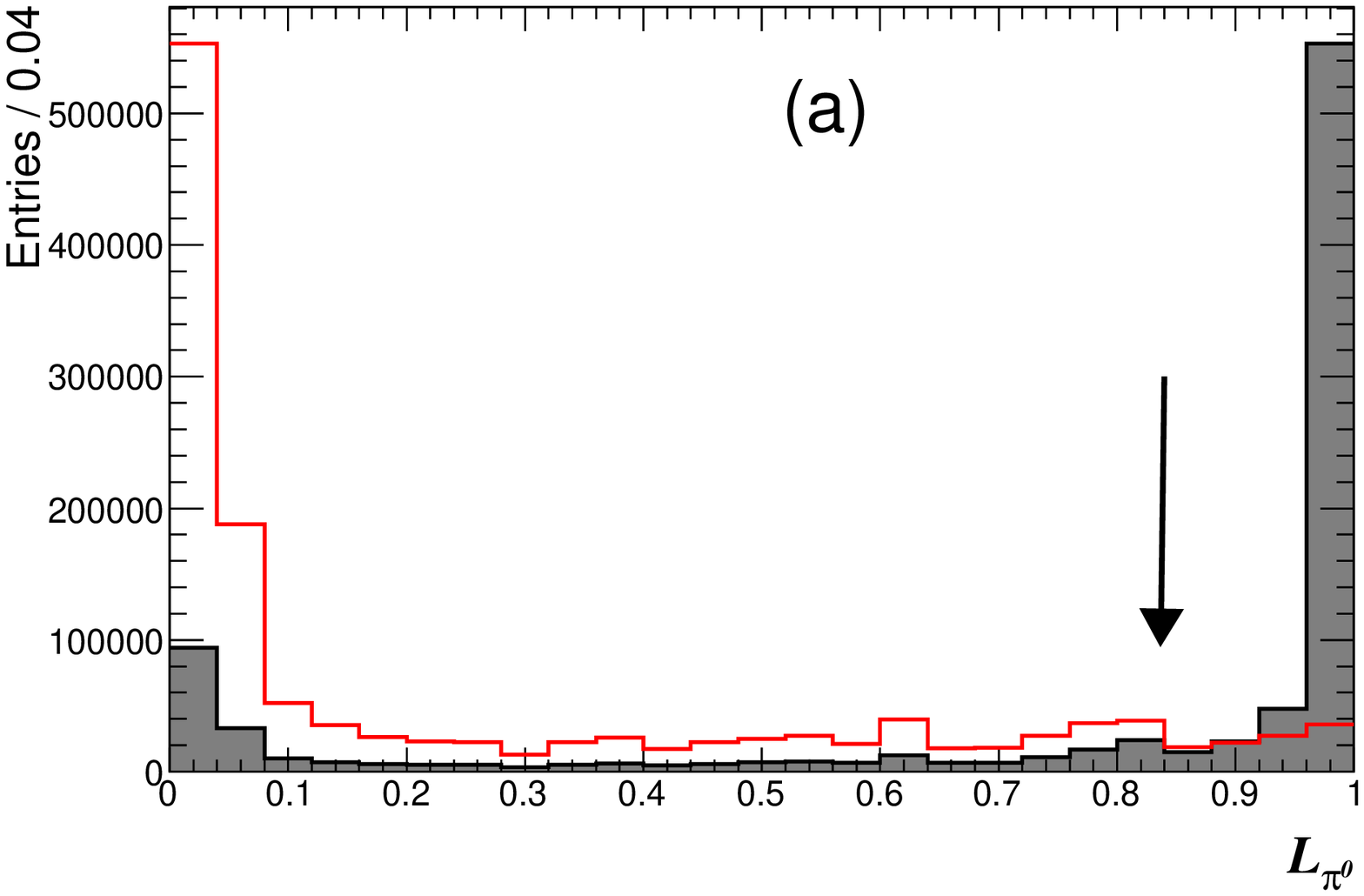}
  \includegraphics[width=\linewidth,clip=true]{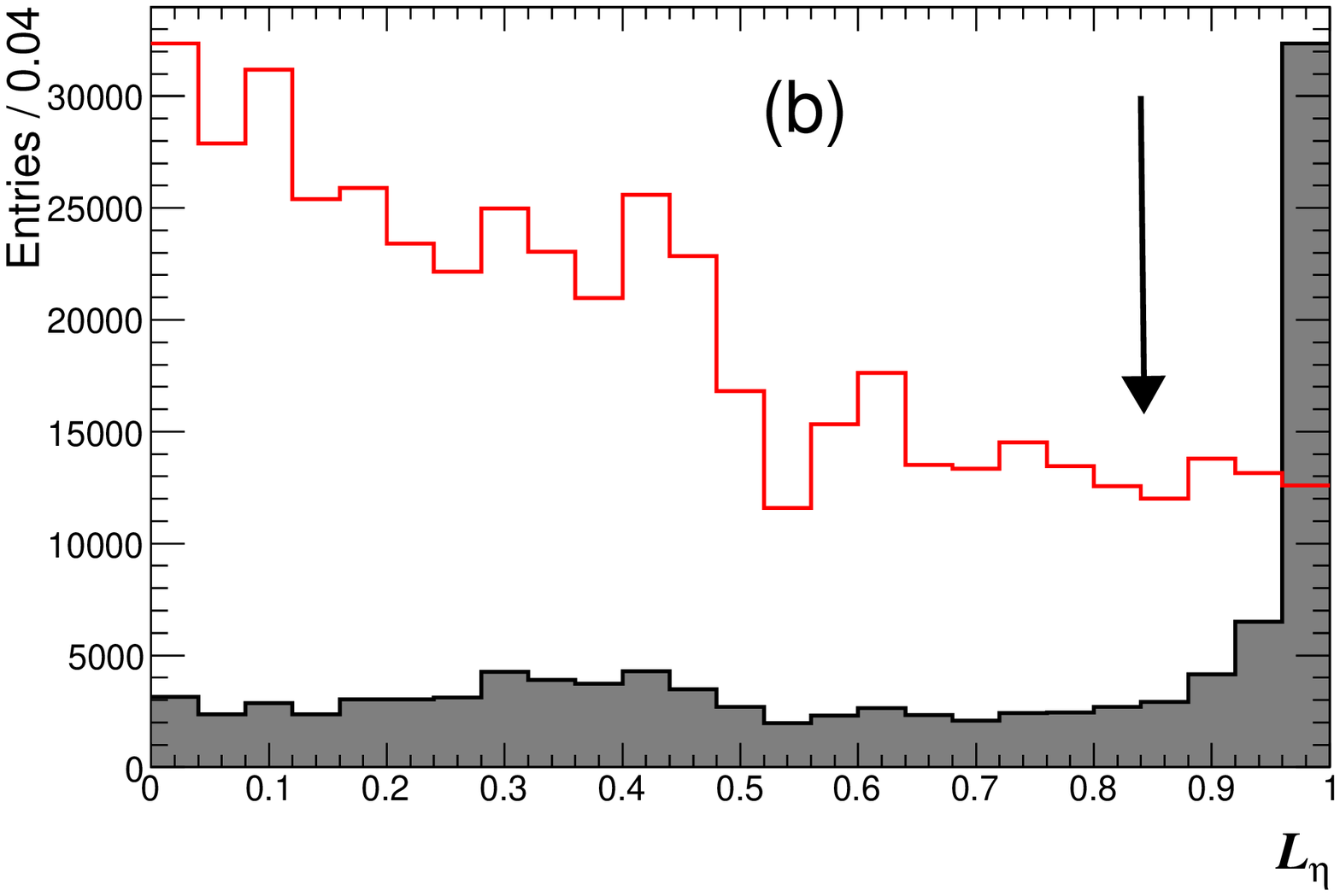}

  \caption
  {
    \label{fig:LR}
    (a)  \piz likelihood  ratio for  \B  candidate photons  in simulated  signal
    events (open  histogram) and  simulated continuum background  events (shaded
    histogram) where the photon is required to originate from a \piz decay.  (b)
    $\eta$ likelihood ratio for \B  candidate photons in simulated signal events
    (open   histogram)  and  simulated   continuum  background   events  (shaded
    histogram)  that are  required to  originate from  an $\eta$  decay.  Events
    where both  \B candidate photon likelihood  ratios are less  than $0.84$ are
    selected, as denoted by the arrows.  
  }

\end{figure}

The dominant  source of backgrounds are  photons produced from  high energy \piz
and  $\eta$ decays in  continuum events.   These events  are suppressed  using a
likelihood ratio  rejection technique.  Each  \B candidate photon  is separately
combined with all other photons in  the event and the invariant mass, $m_{\gamma
\gamma'}$, and energy of the  other photon, $E_{\gamma'}$, are used to calculate
a likelihood ratio given by
\begin{equation}\label{eq:lrRatio}
  L_{i} = \frac{P_{i}(m_{\gamma\gamma'},E_{\gamma'})}
              {P_{\rm sig}(m_{\gamma\gamma'},E_{\gamma'}) +
	       P_{i}(m_{\gamma\gamma'},E_{\gamma'})}.
\end{equation}
In  this equation,  $i$ is  a label  for \piz  or $\eta$,  and $P$  represents a
two-dimensional  probability  density function  (PDF).   For  each \B  candidate
photon  the pairing  that gives  the largest  value of  the likelihood  ratio is
assigned.  The signal PDF, $P_{\rm  sig}$, is constructed using simulated \BzBzb
events containing a \bgg\ decay where all \B candidate photon pairings are used.
The PDF  for a \piz or $\eta$,  $P_i$, is constructed from  simulated $\epem \to
\qqbar$ and  \epem\to\tautau events.   The \B candidate  photon in this  case is
required to  be produced  from a \piz  or $\eta$  decay, while the  other photon
daughter is required to be reconstructed  in the calorimeter.  The energy of the
other photon and the  invariant mass of the pair are then  used to construct the
\piz and $\eta$ PDFs.  A likelihood  ratio near 1.0 (0.0) is consistent with the
\B  candidate   photon  originating  from   a  \piz  or  $\eta$   (signal  $B$).
Figure~\ref{fig:LR}  shows the \LRpiz\  and \LReta\  distributions for  \bgg\ MC
events and for \B candidate photons  from \piz and $\eta$ decays in MC continuum
background events.

For high energy \piz decays with $E_{\piz} \gtrsim 2$ \gev, the daughter photons
may not  be separated enough  in the EMC  to be resolved individually.   In this
case the photon clusters  are said to be ``merged''.  A merged  \piz can mimic a
\B candidate photon because the cluster  will have the full energy of the parent
\piz and will have no associated track.  At a given energy, the second moment of
the energy distribution around the center of the cluster will be different for a
photon and a merged \piz.  This allows for the construction of a quantity called
the  merged \piz  consistency  based on  the  energy and  second  moment of  the
cluster.  This variable compares the two inputs against known distributions from
photons as  well as  merged $\piz$  decays to estimate  the likelihood  that the
cluster originates from  either source.  The \piz energy  range that contributes
\B candidate photons  in this analysis begins at about  2~\gev and extends above
6~\gev.  To reduce this source of merged \piz background, we select \B candidate
photons whose merged \piz consistency is compatible with that of a photon.

\begin{figure}

  \includegraphics[width=\linewidth,clip=true]{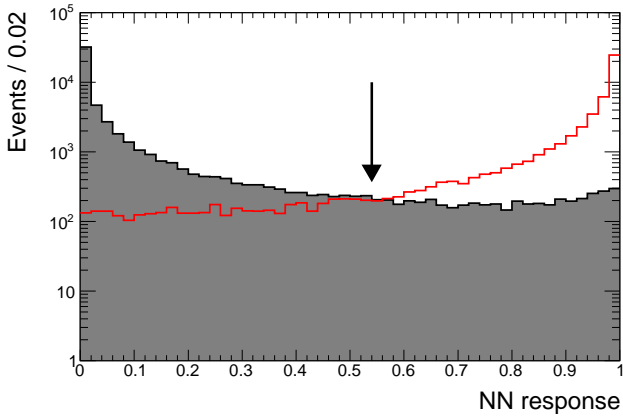}

 \caption
 {
   \label{fig:NNoutput}
   Output of  the neural network on  the validation samples  of simulated signal
   (open histogram)  and background (shaded histogram)  events.  Selected events
   have a NN response greater than 0.54 as denoted by the arrow.
 }

\end{figure}

In  the  \FourS  CM frame,  the  \B  mesons  are  produced  nearly at  rest  and
subsequently decay isotropically, whereas events produced in continuum processes
are  typically collimated in  jets along  the \qqbar  axis.  This  difference in
event shape  is exploited to  further separate signal from  continuum background
events using a neural network  (NN) multivariate classifier.  The NN utilizes 19
input variables that characterize  event level features whose distributions show
separation power between signal and continuum background events.
The inputs  include the  minimum distance between  each \B candidate  photon EMC
cluster and  all other  EMC clusters  in the event,  the polar  angle of  the \B
candidate  momentum  in   the  lab  frame,  the  number   of  neutral  particles
reconstructed, the  number of tracks  in the event,  the missing energy  and the
total  transverse momentum.  We  also include  quantities that  characterize the
rest-of-the-event  (ROE), calculated  using all  reconstructed particles  in the
event except  one or both \B candidate  photons.  These are: the  polar angle of
the event  thrust axis when  either \B candidate  photon is removed;  the first,
second, and third angular moments of the event when the \B candidate photon with
the larger  energy in the  lab frame is  removed; and the second  angular moment
with respect to the thrust axis of  the event when both \B candidate photons are
removed.   Additionally, $R_2$, the  total sphericity,  and the  sphericity with
both  \B  candidate photons  removed  are used;  all  energies  and momenta  are
calculated in the CM frame.
We train and validate the NN using independent samples of \bgg\ and continuum MC
events.   The training  samples are  constructed  by first  applying the  photon
quality,  event pile-up,  and  merged \piz  selections  to the  MC events.   The
surviving events  are randomly  divided into  one set for  training and  one for
validation.  Each set contains 45,200 events  where half are \bgg\ MC events and
the other half  are continuum MC events whose  composition of \epem\to\qqbar and
\epem\to\tautau events is scaled to  the luminosity of the on-resonance data for
each component.  The parameters of the NN are tuned to achieve the highest level
of  background  rejection  while  avoiding  overtraining  the  classifier.   The
validation sample is then used to  verify its performance.  The NN response is a
value     between    0.0     (background-like)     and    1.0     (signal-like).
Figure~\ref{fig:NNoutput} shows  the NN response for the  validation samples for
\bgg\ and continuum background MC events.

The selection  criteria for \LRpiz, \LReta,  the NN response, and  the number of
tracks are optimized using \bgg\  MC events and on-resonance sideband data.  The
number of background events is estimated by extrapolating the sideband data into
the signal  region.  The  optimization proceeds by  iterating over the  space of
each    variable   individually    until   we    find   a    maximum    of   the
figure-of-merit~\cite{punzi} given by the  expression $\varepsilon_{\rm sig} / (
3/2 + \sqrt{B} )$.  In  this equation, $\varepsilon_{\rm sig}$ is the efficiency
of the event selection derived from \bgg\ MC and $B$ is the number of background
events.  The  iterative process continues  to cycle through all  variables until
the selection  values converge.  We find  the optimum values to  be: $\LRpiz <\,
0.84$, $\LReta  <\, 0.84$, NN  response greater than  or equal to 0.54,  and the
number of  tracks to be  greater than two.   The optimum selection  criteria are
found  to have  an overall  efficiency  of $26.7\%$  on a  collection of  $1.96$
million  simulated \bgg\ events,  while rejecting  about $99.9\%$  of background
events.

\subsection{Exclusive \boldmath{$B$} Decay Backgrounds}
\label{sec:Bkg}
Backgrounds from \B  decays that may peak in the \mes  and \DeltaE signal region
are studied with large samples of  simulated events.  Twelve \B decay modes were
identified  as  potential background  sources,  and  exclusive  MC samples  were
generated for  each mode.  The optimized  event selection is applied  to each of
these  samples  and  the  estimated  number of  background  events  expected  in
on-resonance    data    are     determined    from    the    latest    branching
fractions~\cite{PDG2010}.   After  scaling the  yields  of  these  modes to  the
luminosity  of the on-resonance  data, it  is estimated  that they  contribute a
total of $1.18 \pm 0.22$ background events to the signal region.  This number is
comparable  to  the expected  number  of signal  events  predicted  from the  SM
branching  fraction  ($\sim  4$  events).   The  modes  expected  to  contribute
significantly are $\Bz \to \piz \piz$, $\Bz \to \piz \eta$, $\Bz \to \eta \eta$,
and $\Bz \to \omega \gamma$.  The  \mes distributions of these modes peak at the
same value  as true  signal events,  while the \DeltaE  distributions peak  at a
value less  than zero.   This difference in  shape of the  \DeltaE distributions
between \bgg\ decays and these  ``peaking'' background \B decays is exploited by
adding a component that describes them to the maximum likelihood function.

\section{Maximum Likelihood Fit}
\label{sec:MLFit}
The signal yield is extracted  using a two-dimensional unbinned extended maximum
likelihood (ML)  fit in the  region $\mes >  5.2$ \gevcc and $-0.5  \leq \DeltaE
\leq 0.5$ \gev.  The likelihood function for a sample of $N$ events with signal,
continuum, and peaking \BB background components is given by
\begin{equation}\label{eq:likelihoodFunc}
  {\cal L} = \exp \left( - \sum_{i=1}^{3} n_i \right)
                  \left[ 
		    \prod_{j=1}^{N}
		      \left( 
		        \sum_{i=1}^{3} n_i {\cal P}_i( \vec{x}_j; \vec{\alpha}_i)
		      \right)
		  \right],
\end{equation}
where $i$ in this  equation is an index for the three  components in the fit and
$n_i$ is  the event  yield for  each.  Since the  correlations between  \mes and
\DeltaE are found to be small,  the signal and continuum background PDFs, ${\cal
P}_i$, are each defined as a  product of one-dimensional PDFs in the observables
$x_{j}   \in   \{\mes,   \DeltaE\}$,   with  parameters   $\vec{\alpha}_i$.    A
two-dimensional histogram PDF is used for the peaking background component.

The signal PDF  shapes for \mes and \DeltaE are  determined from simulated \bgg\
events.    The  \mes   distribution   is  parameterized   by   a  Crystal   Ball
function~\cite{Gaiser}, and the \DeltaE shape is parameterized by a double-sided
modified Gaussian with tail parameters given by Equation~\eqref{eq:cruijff}.  In
the ML  fit, the signal  PDF parameters are  fixed to the  MC-determined values.
All  fixed  signal  parameters  are  later varied  to  evaluate  the  systematic
uncertainty that this choice of parameterization has on the signal yield.

The  continuum  background  \mes  distribution  is  parameterized  by  an  ARGUS
shape~\cite{ARGUS},  while the \DeltaE  distribution is  fit with  a first-order
polynomial.  The endpoint of the ARGUS  function is fixed to the kinematic limit
for \B decays ($5.29$ \gevcc), while  all other parameters are allowed to float.
The  PDF for  the  peaking  background component  is  parameterized using  large
samples  of simulated  exclusive  \B decays  in  the form  of a  two-dimensional
histogram PDF in  \mes and \DeltaE.  Both the shape and  yield of this component
are fixed in the ML fit.  The yield is fixed to $3.13 \pm 0.54$ events, which is
the predicted number in the fit region determined from the exclusive MC studies.
The fixed  peaking background PDF shape  and yield are later  varied to evaluate
the systematic uncertainty on the signal yield.

The  fit is  validated on  an ensemble  of prototype  datasets whose  signal and
background content and shape are as  expected in the on-resonance data.  For the
signal  content,  one-half,  one,  and  two  times  the  expected  SM  branching
fractions~\cite{BB} are used  to populate the prototype datasets.   Two types of
datasets are  constructed: one where both  the signal and  background events are
generated by randomly  sampling from their respective PDFs,  and the other where
the background events are generated from a random sampling of the background PDF
while the signal events are embedded directly from the simulated signal dataset.
The results of the validation show that the fit returns the correct signal yield
and does not introduce a bias.

\begin{figure}

  \includegraphics[width=\linewidth,clip=true]{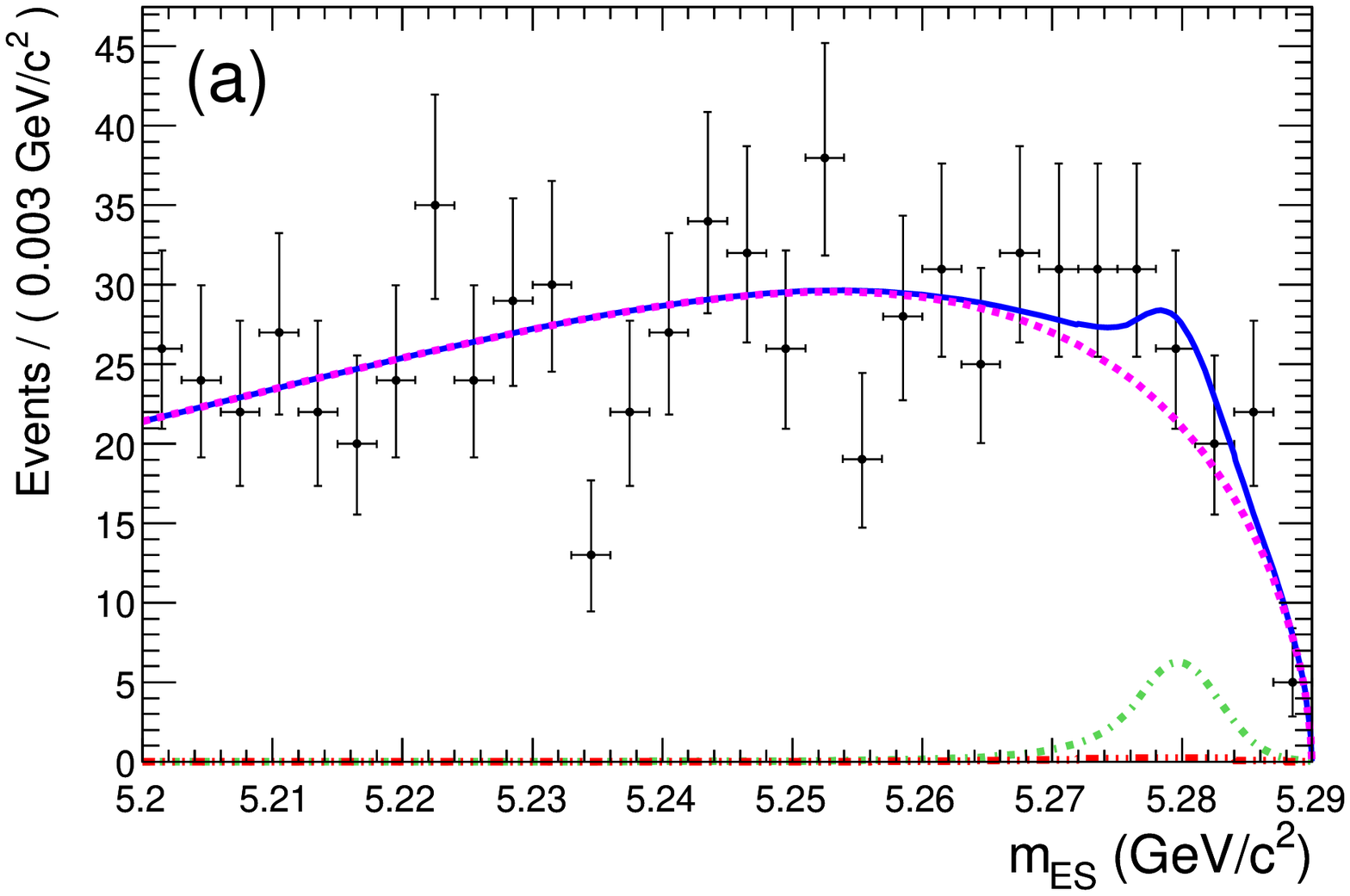}
  \includegraphics[width=\linewidth,clip=true]{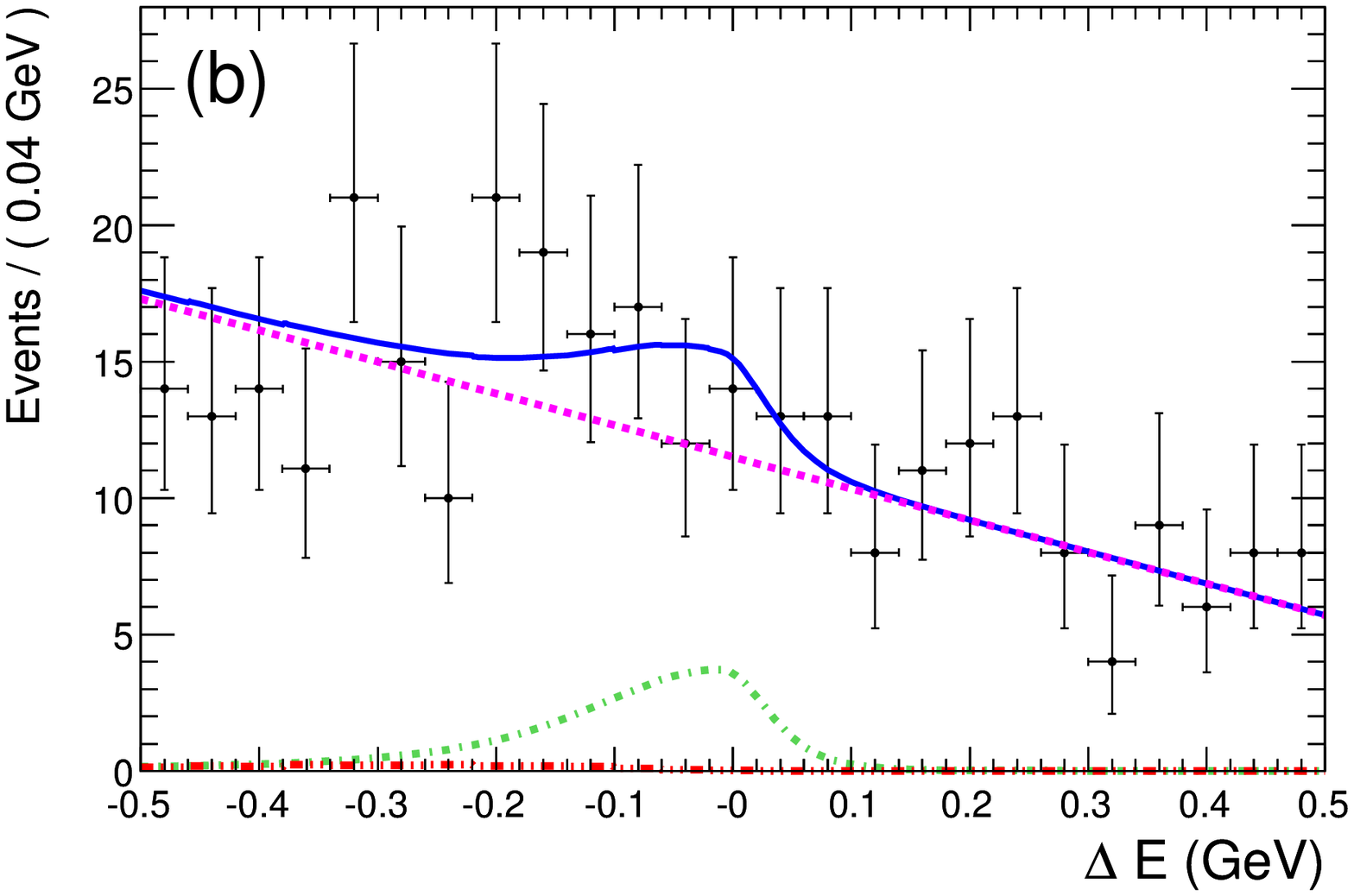}

  \caption
  {
    \label{fig:fit}
    Projections of the ML fit onto  \mes and \DeltaE.  (a) The projection of the
    \mes component when the range of  \DeltaE has been restricted to $-0.30 \leq
    \DeltaE \leq  0.13$ \gev.  (b)  The \DeltaE projection  of the fit  when the
    range  of \mes  has been  restricted to  $\mes >  5.27$ \gevcc.   The points
    represent the on-resonance data.  The  solid curve represents the total PDF,
    the dashed curve is the continuum background component, the dot-dashed curve
    is the signal component, and the long-dashed curve is the peaking background
    component.  With an  expected yield of approximately one  event, the peaking
    background component is nearly indistinguishable from the $x$-axis.  
  }

\end{figure}

The  on-resonance \FourS  data contains  1679 events  after the  optimized event
selection criteria  are applied.  We  perform the ML  fit to extract  the signal
yield  and  find  $N_{\rm  sig}  = 21^{+13}_{-12}$  events  corresponding  to  a
statistical  significance of $1.9\,  \sigma$.  The  significance is  computed as
$\sqrt{2 \cdot \Delta  {\rm ln} {\cal L}}$, where $\Delta {\rm  ln} {\cal L}$ is
the difference in  the log-likelihood between the best  fit to on-resonance data
and a fit  where the signal yield is fixed  to zero.  Figure~\ref{fig:fit} shows
projections of the PDF components from the ML fit.  For the \mes projection, the
range of  \DeltaE has been  restricted to $-0.30  \leq \DeltaE \leq  0.13$ \gev.
For the  \DeltaE projection, the  range of \mes  is restricted to $\mes  > 5.27$
\gevcc.

\section{Systematic Uncertainties}
\label{sec:Systematics}
Systematic  errors that  affect the  calculation of  the branching  fraction are
investigated and  include uncertainties  on the number  of \BzBzb events  in the
dataset,  signal efficiency,  and the  signal yield  from the  fit.  Differences
between data and MC can lead to  an error on the derived signal efficiency.  The
identified  sources  that  can  lead  to this  error  include  uncertainties  in
tracking,  track multiplicity,  photon reconstruction,  the \LRpiz\  and \LReta\
requirements, and  the NN selection.  The  uncertainties in the  modeling of the
signal and \BB  background shapes in the maximum likelihood  fit can also affect
the uncertainty on the signal yield.

The systematic  error associated with  counting the number  of \BB pairs  in the
dataset is $1.1\%$.   The number of \BzBzb pairs is  obtained by multiplying the
\FourS\to\BzBzb    branching   fraction,    equal   to    $(48.4    \pm   0.6)\,
\%$~\cite{PDG2010}, with the total number  of \BB pairs, from which a systematic
error of  $1.7\, \%$  is assigned.   A study of  the track  finding inefficiency
results  in an  assignment  of a  $0.2  \, \%$  systematic  uncertainty for  the
selection of events  with at least three reconstructed  tracks.  The uncertainty
in the signal efficiency due to  the requirement of at least three reconstructed
charged  tracks  is  estimated  to  be  3.4\%,  including  components  for  both
generator-level  simulation  errors   and  detector-associated  data  versus  MC
differences.  The efficiencies in data  and MC for detecting high-energy photons
which pass the  selection criteria were compared using a sample  of $e^+ e^- \to
\mu^+ \mu^- \gamma$ events with the  photon energy in the CM frame restricted to
be consistent  with the energy of  \bgg\ photons.  No  significant difference is
observed.  An uncertainty of 2\% per  photon is assigned to account for possible
data  versus MC differences  due to  the required  minimum distance  between the
candidate-photon cluster and all other  clusters, based on a study that embedded
high-energy photons  in both  data and MC  events.  We combine  this uncertainty
linearly  for both  photons in  \bgg\ and  assign an  overall  photon efficiency
systematic of 4\%.  The cluster time selection is compared in data and MC and we
assign  a  systematic  uncertainty  of  $0.7\%$  for  each  \B  candidate.   The
systematic uncertainty due to the \piz and $\eta$ likelihood ratios is estimated
to be 1.0\% for each, based on a study that embedded signal-like photons in data
and MC events  in which one $B$  meson was fully reconstructed in  the decay $\B
\to D \pi$.  The signal efficiencies  for the \LRpiz\ and \LReta\ selections for
data and MC  are calculated by pairing the embedded  signal-like photon with all
other photons  in these  events that are  not associated with  the reconstructed
$B$.  The  systematic uncertainty due  to the NN  is estimated by  comparing the
efficiencies  of data  to  MC  in signal-like  events.   Signal-like events  are
selected by applying all event  selection criteria, but reversing either the NN,
the \LRpiz\ or  the \LReta\ selection for one of the  \B candidate photons.  The
efficiencies are then  calculated from events in the fit  region with the signal
region  excluded.   For all  selection  reversal  scenarios,  the ratio  of  the
efficiencies is found to be consistent  with unity and has a typical statistical
error of $3.0\%$, which is taken as the systematic uncertainty.

The systematic  uncertainty on the signal yield  due to the choice  of fit model
has four components.   The first is due  to the fixed signal shape  for the \mes
and \DeltaE PDFs.  To estimate the systematic uncertainty related to this choice
of  parameterizations, the  fixed parameters  are varied  within their  $\pm 1\,
\sigma$ errors, the on-resonance dataset is  refit, and the change in the signal
yield is  calculated for  each parameter.  The  total systematic  uncertainty is
taken to  be the sum in  quadrature of all variations  and is found  to be $0.6$
events.  A comparison between photon  response in $\epem \to \mu^+ \mu^- \gamma$
events in data and  MC, using photons in the energy range  relevant to the decay
\bgg, shows  that the size  of the variation  in the signal shape  parameters is
sufficient to take  into account any systematic effects  from parameterizing the
signal PDF shapes from MC.
The second component is due to the parameterization choice for the signal shape.
While the  PDF used to fit the  \mes distribution replicates the  shape in \bgg\
MC, there is a slight disagreement  between the \DeltaE distribution in \bgg\ MC
and  that   of  the  double-sided   modified  Gaussian  used  to   describe  it,
Eq.~\eqref{eq:cruijff}.  To test how large of an effect this difference may have
on the signal  yield, the \DeltaE distribution is  parameterized using a Crystal
Ball shape~\cite{Gaiser} that  provides a larger discrepancy from  the MC shape.
Ensembles  of  simulated  experiments  are  performed  wherein  each  experiment
consists of an independent dataset  fit first using the \DeltaE parameterization
described   in   Section~\ref{sec:MLFit}   and   then  using   the   alternative
parameterization described  here.  The signal  yields for each fit  are compared
and the average difference is found to  be $0.2$ events which is assigned as the
systematic uncertainty.
The third  component is due  to the choice  of the \DeltaE  continuum background
shape.  Repeating  the fit to data  using a second-order  polynomial for \DeltaE
results in an increase of $1.9$ events  in the signal yield, which we take to be
the systematic error for this component.
The fourth  component is due  to the choice  of shape and normalization  for the
peaking background  PDF, both of which  are fixed in  the ML fit.  The  shape is
fixed from the \mes and \DeltaE distributions of simulated exclusive $B$ decays,
while the yield is fixed to  the expected number of events in on-resonance data.
To estimate  the systematic  uncertainty on the  signal yield, both  the peaking
background yield and  shape are varied.  The uncertainties on  the yields of the
individual  peaking  background  modes  are  added  linearly  to  determine  the
uncertainty on  the total yield  of the peaking  component in the ML  fit.  This
results in  a range  for the peaking  yield between  2.02 and 4.24  events.  The
shape of the peaking PDF is varied by replacing it solely with the shape derived
from $B^{0} \to \pi^{0} \pi^{0}$ MC.  The \DeltaE distribution of this mode most
closely resembles that from \bgg\ MC.  Another ensemble of simulated experiments
were  performed where the  differences in  the signal  yield were  compared when
using these  set of extreme variations  in the peaking component  and those from
the ML fit to data.  We take the maximal change in the signal yield and assign a
conservative systematic  uncertainty of $0.5$  events.  The four  components are
added in  quadrature to give  a systematic uncertainty  for the ML fit  of $2.1$
events corresponding to $9.9\%$.

These results are added in quadrature  to give a total systematic uncertainty on
the  signal yield  of $2.6$  events,  corresponding to  $12.1\%$.  The  separate
contributions    to    the   systematic    uncertainty    are   summarized    in
Table~\ref{tab:systErr}.
\begin{table}

  \caption
  {
    \label{tab:systErr}
    Summary of the systematic uncertainties expressed as a percent of the signal
    yield.
  }

  \vspace{0.25cm}

  \begin{tabular}{lr} 
    \hline

    Source                   &  Uncertainty on $N_{\rm sig}$ (\%) \\
    \hline

    \BzBzb counting          &  1.7 \\  
    Tracking efficiency      &  0.2 \\ 
    Track multiplicity       &  3.4 \\
    Photon efficiency        &  4.0 \\  
    Cluster time             &  0.7 \\  
    \LRpiz\ and \LReta       &  2.8 \\  
    Neural network           &  3.0 \\  
    Fit uncertainty          &  9.9 \\
    \hline
    Sum in quadrature        & 12.1 \\
    \hline
  \end{tabular}

\end{table}

\section{Results}
\label{sec:Results}
The branching fraction is calculated from the measured signal yield using
\begin{equation}\label{eq:bf}
  {\cal B}(\Bz  \to \gamma \gamma)  
               = \frac{N_{\rm sig}}{ \varepsilon_{\rm sig} \cdot 2 \cdot 
		       N_{\BzBzb}},
\end{equation}
where  $N_{\rm  sig}$ is  the  signal yield  from  the  maximum likelihood  fit,
$\varepsilon_{\rm  sig}$  is the  signal  selection  efficiency determined  from
simulated \bgg\ events, and $N_{\BzBzb}$ is the number of neutral \B meson pairs
in the on-resonance dataset. We calculate the branching fraction to be
\begin{equation}\label{eq:bf_result}
  {\cal B}(\Bz \to \gamma \gamma ) = ( 1.7 \pm 1.1 \pm 0.2 ) \times 10^{-7},
\end{equation}
with  a statistical significance  of $1.9\,  \sigma$, where  the first  error is
statistical and the second is systematic.

\begin{figure}

  \includegraphics[width=\linewidth,clip=true]{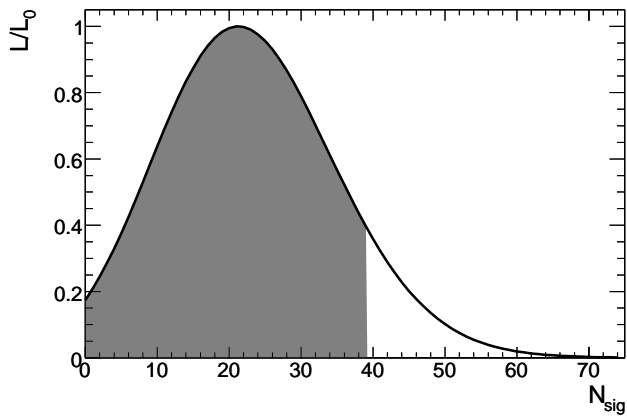}

  \caption
  {
    \label{fig:conv}
    Likelihood  curve from  the maximum  likelihood fit,  as a  function  of the
    signal yield, after  convolution with a Gaussian shape  whose width is equal
    to  the  total systematic  error.   The  shaded  region corresponds  to  the
    integral of the curve up to  90\% of its total area.  The $y$-axis, $L/L_0$,
    is the  ratio of the  likelihood function for  a given $N_{\rm sig}$  to the
    maximum likelihood.  
  }

\end{figure}

The upper limit  at the 90\% CL is obtained by  integrating the likelihood curve
resulting from the ML fit from zero to the value of $N_{\rm sig}$ which contains
$90\%$ of the  area under the curve.  To  incorporate the systematic uncertainty
into the  determination of  the upper limit,  the likelihood curve  is convolved
with a Gaussian  shape whose width is equal to  the total systematic uncertainty
of $2.6$ events.  This yields a value of $N_{\rm sig} = 39$ events corresponding
to an upper limit of
\begin{equation}\label{eq:ul_result}
  {\cal B}(\Bz \to \gamma \gamma) < \result\ \ \ {\rm (90\%\ CL)}.
\end{equation}
Figure~\ref{fig:conv} shows the likelihood  curve from the fit after convolution
with the Gaussian shape.  The shaded  region corresponds to the 90\% integral of
the curve.

This limit  is nearly a  factor of  two below the  best previous upper  limit of
${\cal B}(\bgg)  < 6.2 \times  10^{-7}$ set by Belle~\cite{Belle-Villa},  and is
consistent  with  the SM  branching  fraction.   This  limit may  allow  tighter
constraints on models that incorporate physics beyond the SM.

\section{Acknowledgements}
We are grateful for the 
extraordinary contributions of our \pep2\ colleagues in
achieving the excellent luminosity and machine conditions
that have made this work possible.
The success of this project also relies critically on the 
expertise and dedication of the computing organizations that 
support \babar.
The collaborating institutions wish to thank 
SLAC for its support and the kind hospitality extended to them. 
This work is supported by the
US Department of Energy
and National Science Foundation, the
Natural Sciences and Engineering Research Council (Canada),
the Commissariat \`a l'Energie Atomique and
Institut National de Physique Nucl\'eaire et de Physique des Particules
(France), the
Bundesministerium f\"ur Bildung und Forschung and
Deutsche Forschungsgemeinschaft
(Germany), the
Istituto Nazionale di Fisica Nucleare (Italy),
the Foundation for Fundamental Research on Matter (The Netherlands),
the Research Council of Norway, the
Ministry of Education and Science of the Russian Federation, 
Ministerio de Ciencia e Innovaci\'on (Spain), and the
Science and Technology Facilities Council (United Kingdom).
Individuals have received support from 
the Marie-Curie IEF program (European Union), the A. P. Sloan Foundation (USA) 
and the Binational Science Foundation (USA-Israel).


\begin{thebibliography}{99}

\bibitem{BB}
S.W. Bosch and G. Buchalla,  JHEP 0208:054 (2002).

\bibitem{Aliev-Iltan}
T. M. Aliev and E. O. Iltan, Phys.\ Rev.\ D {\bf 58}, 095014 (1998).

\bibitem{Gemintern-etal}
A. Gemintern, S. Bar-Shalom, and G. Eilam, Phys.\ Rev.\ D {\bf 70}, 035008 (2004).

\bibitem{Belle-Villa}
S. Villa {\it et al.} (Belle Collaboration), Phys.\ Rev.\ D {\bf 73}, 051107 (2006).

\bibitem{Belle-Wicht}
J. Wicht {\it et al.} (Belle Collaboration), Phys.\ Rev.\ Lett.\ {\bf 100}, 121801 (2008).

\bibitem{BABARNIM} 
B.\ Aubert {\em et al.} (\babar\ Collaboration), Nucl. Inst. Meth. A {\bf 479}, 1 (2002).

\bibitem{SLAC-EMC}
M.~Kocian (\babar\ Collaboration), SLAC-PUB-10170 (2002).

\bibitem{evtGen}
D. Lange, Nucl. Inst. Meth. A{\bf 462}, 152 (2001).

\bibitem{jetset}
T. Sjostrand, Comput. Phy. Commun. {\bf 82}, 74 (1994).

\bibitem{geant4}
S. Agostinelli \emph{et al.} (GEANT4 Collaboration), 
Nucl. Inst. Meth. A{\bf 506}, 250 (2003).

\bibitem{PDG2010}
K. Nakamura et al. (Particle Data Group), J. Phys. G {\bf 37}, 075021 (2010). 

\bibitem{fox}
G.C.~Fox and S.~Wolfram, \prl{\bf 41}, 1581 (1978).

\bibitem{Gaiser}
J. E. Gaiser \emph{et al.}, Phys. Rev. D {\bf 34}, 711 (1986).

\bibitem{LAT}
A.~Drescher \emph{et al.} (ARGUS Collaboration), Nucl. Inst. Meth. A {\bf 237}, 464 (1985).

\bibitem{punzi}
G. Punzi, arxiv:physics/0308063v2 (2003).

\bibitem{ARGUS}
H. Albrecht \emph{et al.} (ARGUS Collaboration), Z. Phys. C {\bf 48}, 543 (1990).

\end{thebibliography}
\end{document}